\documentclass{PoS}

\usepackage{graphicx}
\usepackage{amsmath}

\newcommand{\bee}{\begin{equation}}
\newcommand{\ee}{\end{equation}}
\newcommand{\beea}{\begin{eqnarray}}
\newcommand{\eea}{\end{eqnarray}}

\def\Tr{{\rm Tr}}

\title{The complete lowest order chiral Lagrangian from a little box}

\ShortTitle{The complete lowest order chiral Lagrangian from a little box}

\author{\speaker{Thomas DeGrand}\\
        Physics Department, University of Colorado, Boulder CO 80309 USA\\
        E-mail: \email{thomas.degrand@colorado.edu}}

\author{Stefan Schaefer\\
       NIC, DESY, Platanenallee 6, D-15738 Zeuthen, Germany\\
        E-mail: \email{stefan.schaefer@desy.de}}

\abstract{We recently performed a pilot study determining
the parameters of the leading order chiral Lagrangian from 
distributions of the eigenvalues of a quenched  Dirac operator coupled to an imaginary isospin
chemical potential. We complement a quick survey of our recent preprint 
\cite{DeGrand:2007tm} by addressing some points raised during discussions at the
conference.}

\FullConference{The XXV International Symposium on Lattice Field Theory\\
                 July 30 - August 4 2007\\
                 Regensburg, Germany}

\begin{document}

\section{The project, in a nutshell}
Two coupling constants, $F$ and $\Sigma$, parameterize the
leading-order chiral effective Lagrangian
\bee
{\cal L}_{\rm eff} = \frac{F^2}{4} \Tr (\partial_\mu U \partial^\mu U^\dagger)
- \frac{\Sigma}{2} \Tr[ {\cal M}(U+U^\dagger)].
\label{calL}
\ee
A long-standing problem for lattice simulations is to determine them
directly from QCD.
Typically, this is done by fitting a correlation function to some
theoretical formula parameterized by $F$, $\Sigma$ and $m_\pi$.
Usually one has to measure a correlation function of operators at long distance
to be sensitive to  these
parameters. This makes the measurement expensive. In addition, the
simulations do not give $F$ and $\Sigma$ directly; instead,
they give the mass-dependent decay constant and condensate. which depend,
 through the standard formulas of chiral perturbation
theory, on $F$ and $\Sigma$ and on the coefficients of higher-order
terms in ${\cal L}$. So the determination of $F$ and $\Sigma$ from the lattice
 is rather indirect.

A series of papers by Akemann, Damgaard, Heller, Osborn, Splittorf, Svetitsky,
and Toublan\cite{Damgaard:2005ys,Damgaard:2006pu,Damgaard:2006rh,Akemann:2006ru}
 provide an alternative path toward a direct measurement
of $F$ and $\Sigma$. This is done via the properties of eigenvalues
of the Dirac operator, from simulations in the epsilon regime
($m_\pi L\ll 1$ but $\Lambda L \gg 1$ for box length $L$ and $\Lambda$
any nonperturbative parameter of QCD except $m_\pi$). The particular
implementation we used is that of Ref. \cite{Akemann:2006ru}.
An imaginary isospin chemical potential $\mu$ is coupled to a doublet
of quenched quarks. The correlator of the eigenvalues from an ordinary
$\mu=0$ simulation, $\lambda_i$, with the quenched eigenvalues, 
$\lambda_j$,
\bee
\rho_{(1,1)}^{(2)\,conn}(x,y) =
\big \langle \sum_i \delta(x-\lambda_i) \sum_j \delta(y-\lambda_j)\big \rangle
- \big\langle \sum_i \delta(x-\lambda_i)\big \rangle\big \langle \sum_j \delta(y-\lambda_j)\big \rangle  
\ee
is a function of
$\Sigma$ and $F$ which is given by Random Matrix Theory (RMT).
The particular formula is given by
Eq.~(3.49) of Ref.~\cite{Akemann:2006ru},
in terms of the usual rescaled variables $\lambda_i \Sigma V$, $\lambda_j \Sigma V$,
 the rescaled mass
$m_q\Sigma V$,
and the rescaled isospin chemical potential $\delta= \mu F \sqrt{V}$.
$V$ is the volume.

To combine data sets, we integrate the data.
Refs. \cite{Damgaard:2005ys} and \cite{Damgaard:2006pu} suggest using
\bee
C(x,\zeta_{max})=\int_0^{\zeta_{max}} dy \rho(x+y,y) \ .
\label{eq:cx}
\ee
We did one more integral:
\bee
I(X,\zeta_{max})= \int_{-\zeta_{max}}^X C(x,\zeta_{max})dx \ .
\label{eq:ix}
\ee
The first integral $C(x)$ shows a spike at $x=0$ whose width goes
roughly as $\delta^2$. Then $I(X)$ will show a sharp step at $X=0$.

In our numerical simulations we used  overlap 
fermions\cite{Neuberger:1997fp,Neuberger:1998my}.
 They preserve the
full $SU(N_f)\otimes SU(N_f)$ chiral symmetry
(including the anomalous singlet current and the index theorem).
Our data set uses a lattice volume of $12^4$ points.
The lattice spacing $a$, as determined from the Sommer parameter
 $r_0$ \cite{Sommer:1993ce},
is $r_0/a=3.71(5)$.
The pseudoscalar mass is
$am_\pi=0.329(3)$. This is not in the epsilon regime, but experience shows RMT is
robust
enough to work well outside the epsilon regime, and  we are making a
pilot study. In our conversions to physical units we use $r_0=0.5$ fm.
Details of our action are to be found in  Refs.
\cite{DeGrand:2000tf,DeGrand:2004nq,DeGrand:2006ws}. The one new ingredient
is the gauge connection: we used the differentiable hypercubic smeared link of
 Ref. \cite{Hasenfratz:2007rf}.
We collected about 400 thermalized HMC trajectories of unit length
and analyzed 30  lattices in topological charge sector
 $\nu=0$ and 75 $|\nu|=1$ ones. To do the fitting, we took bootstrap
averages of the integrated data.

\begin{figure}
\begin{center}
\includegraphics[width=0.4\textwidth,clip]{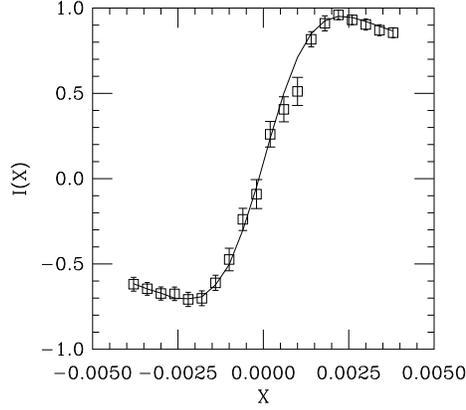}
\end{center}
\caption{Data (squares) and fit from
$|\nu|=1$. A cut $\zeta_{max}=0.07$ is enforced. The best fit values for this bootstrap
sample
are $\Sigma_L V=147$ and $aF_L=0.071$.
\label{fig:exstd}
}
\end{figure}

\begin{figure}
\begin{center}
\includegraphics[width=0.4\textwidth,clip]{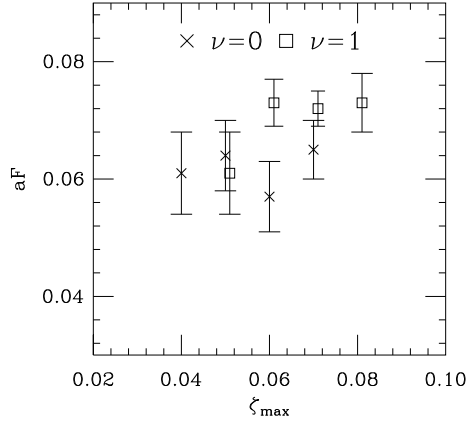}
\end{center}
\caption{Best-fit values of $aF_L$, varying $\zeta_{max}$.
\label{fig:comparezetas}
}
\end{figure}

We reproduce two figures from the paper:
Fig.~\ref{fig:exstd} shows an example of a fit from one of our bootstraps.
Fig.~\ref{fig:comparezetas} shows the dependence of fits on $\zeta_{max}$.
These fits all include only the three lowest (but nonzero)
 eigenvalues in a topological sector.
Smaller  $\zeta_{max}$ means that less data is included in the fit; larger
 $\zeta_{max}$ means that more eigenmodes are needed to saturate
 the correlation function.
It appears that results for this data set do not depend on the choice 
of  $\zeta_{max}$.

In finite volume, the eigenvalue correlator is not a direct function of $\Sigma$ and $F$,
but of the quantities  $\Sigma_L$ and $F_L$ which reflect the finite volume.
For the condensate, this dependence is known:
$\Sigma_L = \rho_\Sigma \Sigma$
where
\bee
\rho_\Sigma = 1 + C_\Sigma \frac{1}{F^2} \Delta(0) +\dots
\ee
with $\Delta(0)$  the contribution to the tadpole graph (propagator
at zero separation) from finite-volume image terms. In the epsilon
regime, $\Delta(0) = -D/\sqrt{V}$
and $D$ depends on the geometry\cite{Gasser:1986vb}. (It is 0.1405 for
hypercubes.)  For $N_f$ flavors, $C_\Sigma = -(N_f^2-1)/N_f$.

A second complication for the condensate is that $\Sigma$ is scheme-dependent. We perform
the conversion from lattice regularization the $\overline{MS}$ using a measurement of the
matching factor from the RI-MOM scheme as an intermediary. Correcting for finite volume
and converting schemes,
we find
\bee
r_0 \Sigma (\overline{MS},\mu=2 \ {\rm GeV})^{1/3} = 0.594(13)
\ee
or
\bee
( \Sigma(\overline{MS},\mu=2 \ {\rm GeV}) )^{1/3} = 234(4) {\rm MeV}
\ee
This is consistent with an analysis of cumulants of eigenvalues, and with other
determinations of the $N_f=2$ 
condensate\cite{McNeile:2005pd,DeGrand:2006nv,Lang:2006ab,Fukaya:2007fb}.

$F$ is also expected to be volume dependent, and the correction is also likely
to be of the form $F_L = \rho_F F$ with 
\bee
\rho_F = 1 + C_F \frac{1}{F^2} \Delta(0) +\dots
\ee
Unfortunately, there is no calculation (yet) of $C_F$. We found
\bee
r_0 F_L = 0.255(13)
\ee
or with $r_0=0.5$ fm,
\bee
F_L = 101(6) {\rm MeV}.
\ee
We expect that $C_F$ is a number on the order 1-2, probably $N_f$ dependent; that would be sufficient
to lower $F$ from $F_L$ by 20 per cent or so and bring it in line with
phenomenological estimates of about 86 MeV \cite{Gasser:1983yg,Colangelo:2003hf}.

So at this point the calculation is incomplete: one needs the finite volume correction.
However, the size of the error is very interesting: while typical large scale
simulations can get $f_\pi$ to a fraction of a per cent, $F$ is harder: for example,
MILC\cite{Aubin:2004fs} quotes
(their Table IV, Fit A, our conversion) $r_1 F =0.131(10)$ or  $F=82(6)$ MeV, compared
to their number for $f_\pi$, which has a
 0.3 per cent error \cite{Bernard:2006wx}). Our quite competitive error on $F$, 
with the expenditure of a tiny amount of computer resources, is what makes this calculation 
worth repeating.

\section{Questions from the conference}

\begin{itemize}
\item
You used fat links. Aren't you worried they will mess something up?
\end{itemize}
On the contrary. Everyone who has tried using fat links for light fermions has only 
good things to say about them, from the point of view of perturbation theory, exceptional
configurations (for Wilson fermions), taste symmetry restoration (for staggered fermions) and
efficiency in computing the overlap action. The one exception was MILC,
who long ago\cite{Bernard:2002pc}
 tried to use APE smearing with $0.45 \times 10$ (!!!) smearing steps
in a calculation of the properties of heavy-light mesons. This
gave big scaling violations in $f_D$: too much smearing decoupled the quarks from the
short distance part of the one gluon exchange potential. We are using hypercubic smearing.
This is even more local than AsqTad -- in fact, you cannot be more local.
Readers might remember that the Alpha collaboration studied static-light properties
with hypercubic links, with great success \cite{Della Morte:2003mn}. 

\begin{itemize}
\item
Why did you measure $F$? Isn't $f_\pi$ more interesting?
\end{itemize}

Actually, what is interesting is the ratio $f_K/f_\pi$, which can be used to determine
the Cabibbo angle\cite{Marciano:2004uf}. But then you have to ask, how do you measure $f_\pi$?
This is not done with simulations at the physical quark masses; instead, one performs
simulations at any light quark masses (heavy enough to be cheaply simulated, light enough
to be in the region of validity of chiral perturbation theory). Then the mass-dependence
of parameters is fit to a chiral perturbation theory formula. The uncertainties 
between $F$, $\Sigma$ and the higher order
terms cancel at the input quark mass values, for the particular observables which are being fit.
However, for other observables (predictions) the larger error on $F$ can be problematic.
That's why this particular measurement scheme -- provided the finite volume
correction can be controlled -- is so interesting.

\begin{itemize}
\item
Isn't there a problem using individual eigenvalues to determine
$F$ and $\Sigma$, rather than using the complete spectral density $\rho(\zeta)$?
\end{itemize}

We don't think so (and anyway, we used formulas which involved integrated
eigenvalue distributions). Ref. \cite{Del Debbio:2005qa} showed that the spectral density
and its moments, and even eigenvalues, renormalize pretty much as one would naively expect.
Notice that at low values of $\zeta$, $\rho(\zeta)$ coincides with the distribution of
the lowest eigenvalue $p_1(\zeta)$ (because that is all that there is)! Akemann and Damgaard
\cite{Akemann:2003tv}
have derived exact expression relating the individual eigenvalue distributions to
integrals over the eigenvalue density and all the spectral correlators. If these are true, 
it is hard to imagine that something different would happen to the individual eigenvalue
distributions and to the integrated ones. (They are also working on a single-eigenvalue
version of the correlator we studied here\cite{Akemann:2007wf}.)

\begin{itemize}
\item
I don't have a chiral action. How much of this project could I do?
\end{itemize}

Doing exactly what we did with fermions which do not recognize the
 index theorem could be tricky, since you have to work in sectors of fixed topology,
and you have to project the nonzero eigenvalues, 
which are complex for a generic non-chiral Wilson-type fermion,
 onto the imaginary axis. This might be more
trouble than it is worth.
You might try looking at quantities in the epsilon regime which average over topology;
the original papers by Hansen  and Leutwyler\cite{hansen} give formulas for the
 pseudoscalar and axial
correlators, which can also be analyzed to get $\Sigma$ and $F$.

\begin{itemize}
\item
Tell me more about the ``Primme'' package.
\end{itemize}

The ``Primme'' package of
McCombs and Stathopolous\cite{primme} is a library for computing eigenvalues of
Hermitian matrices.
(For overlap fermions, that is all you need.)
It implements a variety of algorithms, preconditioning and optimization options
which can be used via a very convenient interface.
It is about a factor of three faster than
our version of the Bunk-Kalkreuter-Simma \cite{ref:eigen} Conjugate Gradient algorithm.

\begin{itemize}
\item
Can you suggest a good book about lattice QCD?
\end{itemize}

Yes \cite{thebook}.

\section{Acknowledgments}
This work was supported by the US Department of Energy.
T.~D. would like to thank Poul Damgaard for his encouragement
and for much valuable advice.
He would also like to thank Kim Splittorf for clarifying correspondence.
We would like to thank Hidenori Fukaya and
Silvia Necco for discussions about finite volume corrections.

\end{document}